\documentclass{article} % For LaTeX2e

\usepackage[utf8]{inputenc}

\usepackage{eurosym}

\usepackage{url}
\usepackage[square,numbers]{natbib}
\usepackage{fontawesome5}
\usepackage{colortbl}
\usepackage{multirow}
\usepackage{lipsum}
\usepackage{amsthm}
\usepackage{mathtools}% Loads amsmathb
\usepackage{xfrac}
\usepackage[export]{adjustbox}
\usepackage{longtable}
\usepackage{makecell}
\usepackage{booktabs}
\usepackage[table,xcdraw]{xcolor}
\usepackage{amssymb}

\providecommand{\doi}[1]{doi: #1}

\definecolor{cai_primary}{HTML}{4C9A99}  % Main CAI color
\definecolor{cai_secondary}{HTML}{307FE2}  % Secondary blue color
\definecolor{cai_accent}{HTML}{1D8348}  % Accent green color
\definecolor{cai_dark}{HTML}{3F4444}  % Dark gray for text
\definecolor{cai_light}{HTML}{F5F5F5}  % Light background color
\definecolor{cai_purple}{HTML}{8A4FFF}  % Purple color for code strings

\usepackage{graphicx}
\usepackage{subcaption}
\usepackage{verbatim}
\usepackage{placeins}
\usepackage{mdframed}
\usepackage{hyperref}
\hypersetup{
    colorlinks=true,
    urlcolor=cai_secondary,
    linkcolor=cai_secondary,    
    filecolor=cai_accent,      
    citecolor=cai_secondary,
}
\usepackage{fancyvrb}
\usepackage{fixltx2e}
\usepackage{bera}
\usepackage{pdfpages}

\usepackage{pgf-umlsd}
\usepackage{tikz} % Required for the CAI architecture diagram
\usepackage[draft=true]{minted}

\usepackage{fancyhdr}
\renewcommand{\headrulewidth}{0.4pt}
\renewcommand{\footrulewidth}{0.4pt}
\renewcommand{\headrule}{\hbox to\headwidth{\color{cai_primary}\leaders\hrule height \headrulewidth\hfill}}
\renewcommand{\footrule}{\hbox to\headwidth{\color{human_color}\leaders\hrule height \footrulewidth\hfill}}

\usepackage{dirtree} % Package for creating directory trees
\usepackage{forest} % Better tree-drawing package
\usepackage{wrapfig} % Added for wrapfigure environment

\usepackage{algorithm} % algorithm style specific (testing)
\usepackage{program} % algorithm style specific (testing)
\usepackage{algorithmic}%testing
\usepackage{listings}
\usepackage{geometry}

\usepackage{caption,setspace}
\DeclareCaptionFormat{listing}{\rule{\dimexpr0.9\columnwidth+17pt\relax}{0.4pt}\par\vskip1pt\textcolor{cai_primary}{\faCode\ #1#2#3}}
\usepackage{titlesec}
\titleformat{\section}
  {\normalfont\Large\bfseries\color{cai_primary}}  % Format
  {\thesection}  % Label
  {1em}  % Separation
  {}  % Before code
  [\titlerule]  % After code

\titleformat{\subsection}
  {\normalfont\large\bfseries\color{human_color}}
  {\thesubsection}
  {1em}
  {}

\titleformat{\subsubsection}
  {\normalfont\normalsize\bfseries\color{cai_dark}}
  {\thesubsubsection}
  {1em}
  {}

\newcounter{code}



\DeclareMathVersion{sans}
\SetSymbolFont{operators}{sans}{OT1}{cmbr}{m}{n}
\SetSymbolFont{letters}  {sans}{OML}{cmbrm}{m}{it}
\SetSymbolFont{symbols}  {sans}{OMS}{cmbrs}{m}{n}

\lstnewenvironment{sflisting}[1][]
  {\lstset{#1}\mathversion{sans}}{}

\usepackage[normalem]{ulem}

\definecolor{grayalias}{HTML}{3F4444}
\definecolor{bluealias}{HTML}{307FE2}
\definecolor{cai_color}{HTML}{4C9A99}  % Adjusted to match CAI color

\definecolor{agentsred}{HTML}{FF6A4C}
\definecolor{agentsorange}{HTML}{F99244}
\definecolor{agentsblue}{HTML}{2D55CC}

\definecolor{agentsred2}{HTML}{993333}
\definecolor{agentsorange2}{HTML}{E67E22}
\definecolor{agentsblue2}{HTML}{2C3E50}

\definecolor{human_color}{HTML}{173C47}  % Darker shade for Humans
\definecolor{speed_color}{HTML}{00BCA2}  % Green for Speedfactor

\definecolor{cai_string}{HTML}{2E8B57}    % Sea green, darker than cai_color
\definecolor{cai_comment}{HTML}{708090}   % Slate gray, complementary to teal
\definecolor{cai_keyword}{HTML}{008080}   % Teal, similar to cai_color
\definecolor{cai_background}{HTML}{F5FFFA} % Mint cream, very light teal tint
\definecolor{cai_identifier}{HTML}{20B2AA} % Light sea green
\definecolor{cai_number}{HTML}{2F4F4F}     % Dark slate gray
\definecolor{cai_frame}{HTML}{4C9A99}      % Same as cai_color for frames

\definecolor{cai_string_muted}{HTML}{3D7A5F}    % Muted sea green
\definecolor{cai_comment_muted}{HTML}{7F8C8D}   % More muted gray
\definecolor{cai_keyword_muted}{HTML}{4C9A99}   % Using CAI color directly
\definecolor{cai_background_muted}{HTML}{F8FBFB} % Very subtle off-white tint
\definecolor{cai_identifier_muted}{HTML}{5F9EA0} % Muted cadet blue
\definecolor{cai_number_muted}{HTML}{45545E}     % Darkened slate gray
\definecolor{cai_frame_muted}{HTML}{4C9A99}      % Same as cai_color for frames

\usepackage{authblk}

\renewcommand\Affilfont{\small\normalfont}
\definecolor{cai_affil_color}{HTML}{3F8984} % Slightly darker variant of cai_color

\makeatletter
\renewcommand\AB@affilsepx{\\\protect\Affilfont}
\let\orig@maketitle\maketitle
\renewcommand{\maketitle}{%
  \orig@maketitle%
  \vspace{-1.5em}%
  {\color{cai_color!30}\hrule height 0.5pt}%
  \vspace{1em}%
}
\makeatother

\title{\LARGE\textcolor{cai_primary}{\textbf{Cybersecurity AI: The Dangerous Gap Between Automation and Autonomy}}}
\author[1]{Víctor Mayoral-Vilches}
\affil[1]{
    {\normalfont\textcolor{cai_color}{\textbf{Alias Robotics}}, Vitoria-Gasteiz, Álava, Spain\\
    {\tt\footnotesize\textcolor{cai_color}{\faEnvelope}~victor@aliasrobotics.com}}
}

{
\makeatletter
\renewcommand\AB@affilsepx{ \quad} % inline separator
\makeatother

}

\makeatletter
\renewcommand\AB@affilnote[1]{}
\makeatother

\begin{document}
%\includepdf[pages=-, fitpaper]{SecDevOps_cover.pdf}

\date{}
\maketitle
%===============================================================================
\vspace{-1em}

\begin{abstract}    
    The cybersecurity industry often confuses ``automated'' and ``autonomous'' AI, creating dangerous misconceptions about system capabilities. Recent milestones like XBOW topping HackerOne's leaderboard \cite{xbow2025top1} showcase impressive progress, yet these systems remain fundamentally semi-autonomous— requiring human oversight. Drawing from robotics principles, where the distinction between automation and autonomy is well-established, I take inspiration from prior work and establish a 6-level taxonomy (Level 0-5) distinguishing automation from autonomy in Cybersecurity AI. Current ``autonomous'' pentesters operate at Level 3-4: they execute complex attack sequences but need human review for edge cases and strategic decisions. True Level 5 autonomy remains aspirational. Organizations deploying mischaracterized ``autonomous'' tools risk reducing oversight precisely when it's most needed, potentially creating new vulnerabilities. The path forward requires precise terminology, transparent capabilities disclosure, and human-AI partnership—not replacement.
\end{abstract}

\section{A Roboticist's Approach to Cybersecurity: Why `Automated' and `Autonomous' Are Not Synonyms}

Automation extends human capability through programmed rules, while autonomy requires the system to exhibit agency – to make choices based on understanding, not just pattern matching. In robotics, we've learned that the leap from automation to autonomy isn't just quantitative (more sensors, more compute) but qualitative, requiring fundamentally different architectures that can model uncertainty, learn from experience, and reason about goals while dealing with unstructured scenarios.

As a roboticist, I've studied \cite{mayoral2025offensive} how the cybersecurity field embraces AI with a familiar mix of excitement and concern. The distinction between \textbf{automated} and \textbf{autonomous} systems – fundamental in robotics – is being dangerously confused in cybersecurity marketing. This isn't mere semantics; it's a critical difference that determines how we design, deploy, and trust these systems.

\begin{wrapfigure}[10]{r}[-0.1\width+.5\columnsep]{6cm}\itshape\large
    {\color{cai_color}
    % Automation is deterministic and predictable. Autonomy requires adaptive behavior and intelligence.
    Only by understanding where automation ends and autonomy begins can we build systems that truly augment human capability while maintaining appropriate safeguards in cybersecurity
    }
\end{wrapfigure}
In robotics, \textbf{automation} refers to systems that execute predefined tasks without human intervention. An automated assembly line robot performs the same weld thousands of times with perfect precision. It's deterministic, predictable, and operates within strict boundaries. This is good engineering – but it's not intelligence.

\textbf{Autonomy} requires something fundamentally different: \textit{adaptive behavior}. An autonomous system must perceive its environment, reason about uncertainty, and adapt its actions to achieve goals in situations its designers never explicitly programmed. The key differentiator is not whether the environment is structured or unstructured – automation can handle complexity too – but whether the system exhibits \textit{intelligent adaptation}.

The cybersecurity domain presents challenges remarkably similar to robotics: dynamic environments, adversarial actors, incomplete information, and the need for creative problem-solving. When vendors brand automated scanners as ``autonomous AI,'' they're not just overselling – they're misrepresenting the fundamental nature of their systems. 

\begin{wrapfigure}[6]{l}[-0.1\width+.5\columnsep]{7cm}\itshape\large
    {\color{cai_color}
    % Automation is deterministic and predictable. Autonomy requires adaptive behavior and intelligence.
    When organizations deploy ``autonomous'' tools that are actually automated, they may reduce human oversight precisely when it's most needed.
    }
\end{wrapfigure}
As established in popular robotics essays \cite{brooks2024laws,brooks2024ai_laws}, the danger isn't in the technology itself but in the expectations gap. The distinction has profound implications for system capabilities, limitations, and appropriate use cases. When organizations deploy ``autonomous'' tools that are actually automated, they may reduce human oversight precisely when it's most needed. In robotics, we've seen this lead to spectacular failures when automated systems encounter scenarios outside their programming. The same risks apply to cybersecurity, where adversaries actively seek to exploit exactly these boundary conditions.

As we'll explore, the path forward requires embracing this distinction, not obscuring it. Only by understanding where automation ends and autonomy begins can we build systems that truly augment human capability while maintaining appropriate safeguards in cybersecurity.

\section{From Automation to Autonomy: Levels of AI in Security}

\begin{table}[!h]
    %\centering
    \small
    \setlength{\tabcolsep}{7pt}
    \begin{tabular}{ccccccl}
        \toprule
        \textcolor{cai_color}{\textbf{Level}} & \textcolor{cai_color}{\textbf{Autonomy Type}} & \textcolor{cai_color}{\textbf{Plan}} & \textcolor{cai_color}{\textbf{Scan}} & \textcolor{cai_color}{\textbf{Exploit}} & \textcolor{cai_color}{\textbf{Mitigate}} & \\
        \midrule
        {\color{cai_color} 0} & {\color{cai_color}\texttt{No tools}} &  {\color{red} \textbf{$\times$}} & {\color{red} \textbf{$\times$}} & {\color{red} \textbf{$\times$}} & {\color{red} \textbf{$\times$}} & {\color{cai_color}\textit{Impossible in practice}} \\
        \midrule
        1 & \texttt{Manual} &  {\color{red} \textbf{$\times$}} & {\color{red} \textbf{$\times$}} & {\color{red} \textbf{$\times$}} & {\color{red} \textbf{$\times$}} & Metasploit  \cite{metasploit} \\
        \midrule
        2 & \texttt{LLM-Assisted} &  {\color{cai_color} \textbf{$\checkmark$}} & {\color{red} \textbf{$\times$}} & {\color{red} \textbf{$\times$}} & {\color{red} \textbf{$\times$}} & \begin{tabular}[l]{@{}p{5.3cm}@{}}PentestGPT \cite{deng2024pentestgpt}\end{tabular} \\
        \midrule
        3 & \texttt{Semi-automated} &  {\color{cai_color} \textbf{$\checkmark$}}  &   {\color{cai_color} \textbf{$\checkmark$}} &  {\color{cai_color} \textbf{$\checkmark$}} & {\color{red} \textbf{$\times$}} & \begin{tabular}[l]{@{}p{5.3cm}@{}}AutoPT \cite{wu2024autopt}, Vulnbot \cite{kong2025vulnbot}\end{tabular} \\
        \midrule
        4 & \texttt{Cybersecurity AIs} & {\color{cai_color} \textbf{$\checkmark$}} & {\color{cai_color} \textbf{$\checkmark$}} & {\color{cai_color} \textbf{$\checkmark$}} & {\color{cai_color} \textbf{$\checkmark$}} & \textcolor{cai_primary}{\textbf{CAI}} \cite{aliasrobotics2025cai} \\
        \midrule
        {\color{cai_color} 5} & {\color{cai_color}\texttt{Autonomous}} & {\color{cai_color} \textbf{$\checkmark$}} & {\color{cai_color} \textbf{$\checkmark$}} & {\color{cai_color} \textbf{$\checkmark$}} & {\color{cai_color} \textbf{$\checkmark$}} & {\color{cai_color}\textit{Aspirational}} \\
        \bottomrule
    \end{tabular}
    \caption{The autonomy levels in cybersecurity, adapted from \cite{aliasrobotics2025cai} and SAE J3016 \cite{sae2021j3016} driving automation levels. I classify cybersecurity autonomy from Level 0 (no tools) to Level 5 (full autonomy). Table outlines capabilities each level allows a system to perform autonomously: \texttt{Planning} (strategizing actions to test/secure systems), \texttt{Scanning} (detecting vulnerabilities), \texttt{Exploiting} (utilizing vulnerabilities), and \texttt{Mitigating} (applying countermeasures).}
    \label{tab:pentesting}
\end{table}

Not all ``AI security tools'' are created equal – there's a spectrum from no automation to true autonomy. Adapting prior art \cite{aliasrobotics2025cai} alongside the well-established SAE J3016 levels of driving automation \cite{sae2021j3016}, I propose six levels of cybersecurity autonomy, from Level 0 (no tools) to Level 5 (autonomous):

\begin{itemize}
\item \textbf{Level 0 – No Tools:} The security professional performs all tasks manually, without any computational tools whatsoever. This parallels SAE Level 0 where the driver performs all driving tasks. While some practitioners proudly claim they operate at this level, performing modern security assessments without any tools is akin to claiming you're a surgeon who refuses to use scalpels – technically conceivable, but professionally negligent. Even using a terminal or text editor constitutes tool usage, making true Level 0 operation virtually impossible in practice.

\item \textbf{Level 1 – Manual Tools:} The professional uses tools like Metasploit\cite{metasploit} or Nmap that provide specific assistance, but \textbf{every decision remains human-driven}. Like adaptive cruise control in vehicles, these tools handle repetitive tasks but require constant human control and decision-making. The tools execute commands but don't suggest what to do next.

\item \textbf{Level 2 – LLM-Assisted:} AI provides planning assistance while humans execute. Systems like \textbf{PentestGPT}\cite{deng2023pentestgpt} can suggest attack strategies and generate code, but the human must actively supervise and execute all actions. This mirrors SAE Level 2 where the vehicle controls steering and acceleration but requires constant human monitoring. The AI is an intelligent assistant, not an independent actor.

\item \textbf{Level 3 – Semi-Automated:} The AI system can execute complete attack sequences in specific, well-defined scenarios. Tools like AutoPT\cite{wu2024autopt} and Vulnbot\cite{kong2025vulnbot} alongside many others operate here – they can autonomously scan, exploit, and report findings, but \textbf{require human intervention} for edge cases, validation, and mitigation strategies. Like SAE Level 3 vehicles that handle highway driving but need the driver ready to take over, these systems work well until they encounter unexpected situations.

\item \textbf{Level 4 – Cybersecurity AIs:} Systems aim to handle the complete security assessment lifecycle in security scenarios with minimum human intervention. \textcolor{cai_primary}{\textbf{CAI}}\cite{aliasrobotics2025cai} represents our current work towards this level, capable of planning, scanning, exploiting, and suggesting mitigations autonomously. However, like SAE Level 4 vehicles limited to specific operational domains, these systems still require human oversight.

\item \textbf{Level 5 – Autonomous:} The aspirational goal where AI handles all cybersecurity tasks in all conditions without human intervention. This would be equivalent to a fully autonomous vehicle that needs no steering wheel. Currently, no such system exists, and given the ethical, legal, and strategic complexities of cybersecurity, Level 5 may remain theoretical – a north star for research rather than a practical goal.
\end{itemize}

Most real systems today sit in the middle of this spectrum – we have \textit{LLM-assisted} and \textit{semi-automated} tools,  while fully autonomous security AI is an aspirational goal. The open-source \textbf{Cybersecurity AI (CAI)} framework \cite{aliasrobotics2025cai,cai2025github}, for instance, aims for Level 4 capabilities by combining multiple specialized agents (for pentesting, bug hunting, blue teaming, etc.) with seamless tool integration and a human supervisor overseeing the AI's choices. CAI can autonomously perform planning, scanning, exploitation, and mitigation in a coordinated fashion, making it one of the first platforms to attempt full-spectrum autonomy. Yet even we describe it as \textit{``semi-autonomous''} in practice, because {strategic human insight is still required in complex scenarios}.

\begin{wrapfigure}[12]{l}[-0.1\width+.5\columnsep]{8cm}\itshape\large
    {\color{cai_color}
    Traditional security automation (think vulnerability scanners or scripted exploits) falls in the lower levels – they execute predefined patterns or searches. In contrast, the new AI agents operate at higher levels by {reasoning about their approach}, dynamically adjusting strategies, and handling a broader range of tasks with minimal direct instruction.
    }
\end{wrapfigure}

This graduated view of autonomy helps clarify \textbf{``how are AI agents different from other automation tools and scanners?''}. Traditional security automation (think vulnerability scanners or scripted exploits) falls in the lower levels – they execute predefined patterns or searches. In contrast, the new AI agents operate at higher levels by {reasoning about their approach}, dynamically adjusting strategies, and handling a broader range of tasks with minimal direct instruction. For example, a legacy web vulnerability scanner might blindly crawl a site and flag anything that matches known signatures. An AI agent, on the other hand, can decide which parts of the application seem promising, {write and run a custom exploit on the fly}, and interpret the outcome, much like a human hacker would – all without needing a step-by-step script for each possibility.

A note on Level 0 – No Tools: While I include it for completeness, any security professional claiming to operate \emph{"without tools"} in 2025 is either being disingenuous or professionally incompetent. The very act of typing commands requires a terminal, viewing results requires a display, and even mental arithmetic could be considered a cognitive tool. Unsurprisingly, there are some renowned hackers who do take pride in claiming they use no tools – often as a badge of honor or demonstration of pure skill. However, such practitioners romanticize the notion of ``pure manual testing,'' perhaps confusing minimal tool usage with no tool usage. The real expertise lies not in avoiding tools, but in selecting the right level of automation for each task – knowing when to rely on AI agents, when to use simple scripts, and when human intuition must guide the process.

\section{The Human-In-The-Loop: Why Oversight Matters}

Despite rapid progress in AI, \textbf{human expertise remains a critical component of effective cybersecurity AI systems}. In fact, many so-called ``autonomous'' tools today rely on humans in less obvious ways. Often, the \textit{``Human-In-The-Loop (HITL)''} is not a software engineer tweaking the model in a lab – it's \textit{the end-user or security analyst who provides the prompts, guidance, and final judgment} on the AI's findings. In other words, the user becomes the human overseer, intentionally or not.

\begin{wrapfigure}[7]{r}[-0.1\width+.5\columnsep]{5cm}\itshape\large
    {\color{cai_color}Even the most ``autonomous'' robotic systems today rely on human oversight in critical ways}
\end{wrapfigure}
Take the example of {XBOW}, the AI-driven penetration testing system that made headlines by topping a bug bounty leaderboard \cite{xbow2025top1,helpnet2025xbow}. The company's blog boldly describes XBOW as a \textit{``fully autonomous AI-driven penetration tester''} that \textbf{``requires no human input''}\cite{xbow2025top1}. Presumably, XBOW was designed to operate much like a human pentester would – it can scan thousands of targets and launch exploits in a matter of hours. \textbf{However, in practice a human team was very much in the loop}. XBOW's creators acknowledge that before any vulnerability reports were submitted to HackerOne, \textit{``our security team reviewed them pre-submission to comply with [the] policy on automated tools''}. In other words, {the AI's work had to be vetted by people} to filter out mistakes and ensure quality. This human review step speaks volumes: even an ``autonomous'' bug hunter needed a human safeguard to verify that its 1,060 automatically generated findings were real issues.

This pattern is common in security. {Automation has long struggled with false positives}, and AI is not immune to that challenge. As one analysis of XBOW's performance put it, \textit{``tools that flag dozens of irrelevant issues often create more work than they save''} – an age-old problem in vulnerability scanning. In robotics, we face similar challenges with sensor noise and perception errors\cite{electronics10222850}. A robot might ``see'' obstacles that aren't there, just as a security scanner might flag benign code as malicious. AI systems are powerful at generating hypotheses (they \textit{``generalize well''}), but \textit{``verifying technical edge cases is a different game entirely''}. For this reason, XBOW's team built additional logic called \textit{validators}: essentially, {automated peer-review checks to confirm each vulnerability the AI finds}. For example, if the AI suspected an XSS flaw, a validator would automatically launch a headless browser to see if the malicious payload truly executes in a live page. In some cases they even invoke a second LLM to double-check the first LLM's result. These measures highlight that {today's ``autonomous'' systems are actually closely monitored and quality-controlled by their creators or users}.

Crucially, \textbf{overstating AI's independence can be dangerous}. Misleading claims might cause organizations to place unwarranted trust in a tool or misunderstand what it really does. As one security researcher cautioned, \textit{``overstating AI capabilities can confuse the public and mislead buyers, especially in high-stakes security contexts.''} In the excitement of breakthrough results, it's important to keep sight of who is really solving the problems. In the case of XBOW, it's \textit{not} that an AI magically became a hacker – it's {a team of human researchers leveraging generative AI to automate parts of their work}. The humans {design the system prompts, build and integrate the attack tools, and guide the decision logic} that the AI follows. In short, \textbf{the AI is a powerful assistant, not an autonomous hacker}. A human developer \textit{``designs the brain,''} and a human analyst is on standby as the \textit{common sense} and safety net. This reality should temper how we discuss ``autonomous'' hacking. In many cases, the innovation is real – AI agents are performing tasks that previously required human labor – but they are doing so {under human supervision and strategy}.

Keeping a human in the loop is not just a formality; it remains essential for several reasons. \textbf{Ethical and legal considerations} demand oversight – an AI that finds a critical vulnerability might need a human decision on how to responsibly disclose or remediate it. This mirrors the robotics principle of ``meaningful human control''\cite{rodriguez2017cybersecurity}, where autonomous systems must remain under human authority for critical decisions. There are also still \textit{hard problems} in security that AI alone hasn't solved. For instance, triaging the impact of a bug or reproducing complex multi-step exploits often requires intuition and context that current AIs lack. As one practitioner noted, \textit{``right now [triage] is still mostly manual and it is a hard problem to automate.''} In summary, \textbf{human judgment provides the contextual understanding, ethical guidance, and final verification that even the best AI tools still need}. The most promising vision for the near future is not replacing humans, but \textbf{partnering} humans with AI – letting the machines grind through rote tasks at superhuman speed, while experts steer the overall effort and handle the nuanced decisions.

\section{Hype vs. Reality: Responsible Advancement}

The excitement around AI in cybersecurity is well-founded – the technology \textbf{is} catching previously missed bugs and accelerating security testing in remarkable ways. Open-source projects like CAI \cite{cai2025github} have already demonstrated \textbf{human-competitive performance} in hacking competitions, even achieving first-place among AI teams in a recent live CTF and ranking in the top 20 against human teams \cite{aliasrobotics2025cai}. \begin{wrapfigure}[8]{l}[-0.1\width+.5\columnsep]{5.5cm}\itshape\large
    {\color{cai_color}AI can democratize security testing, helping smaller organizations access capabilities previously limited to big players}
\end{wrapfigure}In real-world trials, these AI agents have enabled non-experts to find genuine security flaws at rates comparable to veteran researchers. Such results suggest that, used wisely, AI can dramatically improve both the \textbf{efficiency} and \textbf{accessibility} of security testing. It can help level the playing field, allowing smaller organizations (who could never afford a full-time security team or expensive consultants) to regularly audit their systems with AI-driven ``pentesters''. This democratization of security is a positive development – it addresses what some call the \textit{``oligopolistic ecosystem''} of security expertise, where only big players have the resources to secure themselves thoroughly.

However, it's important to keep claims grounded. As I've discussed, \textbf{no current AI is a hack-and-forget magic box}. Achievements like XBOW's are a \textit{team effort}, combining human insight and machine speed. Recent funding rounds – XBOW \$75M Series B \cite{xbow2025seriesb}, Horizon3.ai raising \$100M\cite{techcrunch2025horizon3,securityweek2025horizon3}, Mindgard securing \$8M\cite{lancaster2024mindgard,techeu2024mindgard} – show investor enthusiasm, but also raise concerns about overpromising. The founder of XBOW proudly announced that \textit{``our AI now rivals and surpasses top-tier human hackers''} \cite{xbow2025seriesb} – but in the same breath, the details show the AI was buttressed by human reviewers and months of careful tuning on custom benchmarks. \textbf{We should applaud the progress} – it's genuinely impressive that an AI-centered approach can outperform individual human hackers on certain tasks. Yet we must also \textbf{communicate clearly about how it works}. In the case of XBOW, their Agent (made out of proprietary scaffolding) leverages existing security tools and third party US-based LLMs to build the so called AI Hackers. It's really humans + AI together that achieved the result, not an autonomous AI acting in isolation. By giving credit accurately, we avoid feeding unrealistic fears or expectations in the community.

Moving forward, a \textit{responsible} approach to Cybersecurity AI would include: transparency about capabilities and limitations, continued integration of human oversight, and rigorous evaluation of these tools in diverse real-world conditions \cite{achuthan2024advancing}. Encouragingly, the research community is addressing these. For example, with CAI \cite{aliasrobotics2025cai} our team openly published our benchmarks and results to \textbf{counter any excessive hype or downplaying} by vendors. In fact, we unveil how some LLM providers are downplaying the capabilities of their models for security use cases. This helps everyone understand what AI can and \textbf{cannot} do today. Likewise, investors and companies injecting money into ``AI hackers'' (Horizon3.ai, Mindgard/Mindfort, and others have each raised tens of millions) must realize that \textbf{solving cybersecurity is not just a scaffolding problem} – it also involves modeling, data collection, user experience, community engagement, and handling edge cases that defy full automation. The \textit{future of AI in security is exciting}, but success will come from pairing technological breakthroughs with human intuition.

As one of the researchers actively working on these AI security agents, I remain optimistic. Every month seems to bring improvements – bigger or more specialized models, better tool integration, and lessons learned from live tests. We are just getting started on what AI can do for cybersecurity. But I also believe \textbf{we owe the community clarity, not marketing hyperbole}. The vision of fully autonomous cyber defenses and offenses may well be attainable in the coming years, yet it will require \textbf{steady, evidence-based progress}. In the meantime, the best results will come from human experts and AI \textbf{working hand-in-hand}, each covering the other's blind spots. By staying grounded about what AI does behind the scenes, we can \textit{advance this space responsibly} – celebrating real achievements, avoiding overhyped claims, and ultimately building trust in these powerful new tools.

\section{Discussion: Critical Analysis of Autonomy Claims}

The following table synthesizes key challenges to claims of ``full autonomy'' in cybersecurity AI systems, drawing from established principles in robotics and control theory:

\begin{table}[!h]
    \centering
    \small
    \setlength{\tabcolsep}{8pt}
    \renewcommand{\arraystretch}{1.4}
    \begin{tabular}{p{5cm}p{10cm}}
        \toprule
        \textcolor{cai_color}{\textbf{Claim Category}} & \textcolor{cai_color}{\textbf{Critical Analysis}} \\
        \midrule
        \textbf{Autonomy Definition} & \\
        \hspace{0.3cm}\textit{``Practical'' vs ``Absolute''} & Running for hours on a single goal $\neq$ true autonomy. Robotics requires handling unforeseen contingencies without mandatory human gates. Required human clicks break the closed feedback loop, placing systems at Level 3 (semi-autonomous, refer to Table \ref{tab:autonomy-levels-cybersecurity}) at most. \\
        \midrule
        \textbf{Human Involvement} & \\
        \hspace{0.3cm}\textit{Oversight as HITL} & Asynchronous review still constitutes Human-In-The-Loop. Any veto point makes humans functional elements of the decision surface, disqualifying Level 5 autonomy claims. \\
        \hspace{0.3cm}\textit{Validators as Props} & Automated validators (browsers, LLM checks) are scripted unit tests, not strategic reasoning. They reduce false positives but cannot arbitrate novel edge cases—strengthening automation, not proving autonomy. \\
        \midrule
        \textbf{Environmental Complexity} & \\
        \hspace{0.3cm}\textit{Digital Terrain Dynamics} & Cyberspace changes hourly (code pushes, zero-days, countermeasures), creating decision-latency problems comparable to physical robotics. Claims that autonomy ``arrives sooner in cyberspace'' remain unproven. \\
        \midrule
        \textbf{Transparency \& Trust} & \\
        \hspace{0.3cm}\textit{Evidence Gap} & Public releases show redacted walkthroughs omitting raw payloads and reasoning steps. True transparency requires complete and reproducible logs, agent thought processes, and validator outputs (e.g. as CAI \cite{cai2025github} provides). \\
        \hspace{0.3cm}\textit{Marketing vs Reality} & Advertising ``fully autonomous pentesting'' while admitting human controls risks repeating self-driving car over-promises. Clear definitions protect customers and field credibility. \\
        \bottomrule
    \end{tabular}
    \caption{Critical examination of autonomy claims in cybersecurity AI. Each category highlights gaps between marketing language and technical reality, emphasizing the importance of precise terminology when classifying AI capabilities.}
    \label{tab:autonomy-levels-cybersecurity}
\end{table}

\vspace{0.5em}
These distinctions matter beyond academic pedantry. \textbf{When organizations deploy ``autonomous'' tools that are actually automated, they may reduce human oversight precisely when it's most needed—potentially creating new vulnerabilities rather than solving them}. The path forward requires embracing this nuance, not obscuring it for marketing advantage.

\section{Conclusion}

\begin{wrapfigure}[7]{r}[-0.1\width+.5\columnsep]{5.5cm}\itshape\large
    {\color{cai_color}The distinction between automated and autonomous cybersecurity AI is not academic pedantry—it's a critical safety issue.}
\end{wrapfigure}
The distinction between automated and autonomous cybersecurity AI is not academic pedantry—it's a critical safety issue. Today's most advanced systems like XBOW and CAI \cite{cai2025github} operate at Levels 3-4: impressive automation with strategic human oversight, not true autonomy. They find real vulnerabilities faster than humans alone, democratizing security testing for organizations of all sizes. But they remain tools requiring human judgment for validation, ethics, and edge cases.

The danger lies in the marketing gap. When vendors claim ``full autonomy'' while quietly maintaining human review loops, organizations may dangerously reduce oversight. As shown in Table \ref{tab:autonomy-levels-cybersecurity}, even our most sophisticated AI pentesters need human validators to prevent false positives and ensure responsible disclosure.

In summary, \textbf{``automated vs autonomous'' in cybersecurity is not a binary switch but a continuum} partly depicted in Table \ref{tab:autonomy-levels-cybersecurity}. The path forward is clear: embrace the human-AI partnership model. Let machines handle the repetitive scanning and initial exploitation while humans provide strategic direction and ethical boundaries. Demand transparency from vendors about actual capabilities. Use precise terminology—Level 3 semi-automated, not ``fully autonomous.'' Most importantly, maintain human oversight precisely because these tools are powerful, not despite it. The future of cybersecurity AI lies not in replacing human expertise but in amplifying it through honest, responsible advancement.

\section{Acknowledgements}

This research has been partly funded by the European Innovation Council (EIC) as part of the accelerator project ``RIS'' (GA 101161136) - HORIZON-EIC-2023-ACCELERATOR-01 call. The author thanks Endika Gil-Uriarte for his review on the document and suggestions.

% Bibliography will be added here
\bibliography{bibliography}

\end{document}